\documentclass[a4paper,12pt]{article}
\usepackage{amssymb,amsmath,graphicx,subfigure,epsfig}
\usepackage{amsfonts}
\usepackage{color}
\usepackage{epsfig}
\usepackage{amsxtra}
\usepackage{amstext}
\usepackage{latexsym}

\title{A Generalized Master Equation for Janus Faced
Coherent States}
\author{K. V. S. Shiv Chaitanya$^{1}$~~and~~V. Srinivasan$^{2}$\\
$^{1}$The Institute of Mathematical Sciences,\\CIT Campus, Taramani, Chennai, India.\\
$^{2}$Department of Theoretical Physics, University of Madras,\\Guindy Campus, Guindy, Chennai, India.}

\begin{document}

\maketitle
\begin{abstract}
A general master equation for eigenstates of Janus Faced commutation relations 
is constructed.
\end{abstract}

\section{Introduction}
In the paper \cite{vsc}  
it is shown that coherent states occur in pairs. The commutation relations 
$[{\cal F},{\cal G}^\dagger]=1$ permits two eigenstates, 
one for  $ {\cal F}$ and the other for $ {\cal G}$. Given $ {\cal G}\vert v\rangle_i=0$
then the eigenstate for $ {\cal G}$ is defined as $\vert g\rangle_i= exp[g{\cal F}^\dagger]\vert v\rangle_i$
with corresponding eigenvalue $g$ and the eigenstate of $ {\cal F}$ is defined as
${\cal F}\vert f\rangle_i= f\vert f\rangle_i$ with eigenvalue $f$.
It was further shown that the Caves-Schumaker state and pair coherent state 
is one such pair, similarly Yuen and Cat states form another pair. That the
Caves-Schumaker state and pair coherent state can be realized
as short term and steady state solutions of master equation was
shown in the seminal paper on pair coherent states by G.S. Agarwal \cite{ag}.
In this short note we propose a general master equation whose 
short term and steady state solutions are the eigenstates of $ {\cal F}$ and $ {\cal G}^\dagger$. 
A general master equation in terms of Janus Faced pair which is given by
\begin{eqnarray}
\frac{\partial}{\partial t}\rho &=&
-iG\left({\cal F}\rho-\rho {\cal F}+{\cal F}^{\dagger}\rho-\rho {\cal F}^{\dagger}\right)
+\frac{\kappa}{2}
\left(2{\cal F}\rho {\cal F}^{\dagger}
- {\cal F}^{\dagger}{\cal F}\rho-\rho {\cal F}^{\dagger}{\cal F}\right).
\label{jf}
\end{eqnarray}
To solve this equation we use thermo field dynamics (TFD) \cite{[vs1]},
a detailed description of TFD is given below,
In TFD, the averages of operators with respect to $\rho$ reduces to a scalar product.
\begin{equation}
\langle A\rangle = Tr A\rho = \langle\rho^{1-\alpha}\vert A\vert \rho^\alpha\rangle.
\end{equation}
where $\vert \rho^\alpha\rangle,  1/2  \le 
\alpha \le 1$ is a state vector in the extended Hilbert space ${\cal H}\otimes {\cal 
H}^*$ is given by
\begin{equation}
\vert \rho^\alpha\rangle = \hat{\rho}^\alpha\vert I\rangle,
\end{equation}
here $\hat{\rho}^\alpha = \rho^\alpha \otimes I,$
and $\vert I\rangle$ is 
\begin{equation}
\vert I\rangle=\sum \vert N\rangle\langle N\vert = 
\sum \vert N\rangle\otimes\vert N\rangle \equiv \sum \vert N,N\rangle,\label{i}
\end{equation}
in terms of a complete orthonormal set  $\vert N\rangle$  in  ${\cal  H}$.  
 It has been used by S. Chaturvedi and V. Srinivasan in ref \cite{[vs3],[vs4]}
for systems with dissipative dynamics in a Kerr medium.
In this representation, for any hermitian operator $A$, one has 
\begin{equation}
\langle A\rangle = Tr(A\rho) = \langle A\vert\rho\rangle.
\end{equation} 
By choosing the number state $\vert n\rangle$
for  $\vert N\rangle$
 and introducing the creation and  annihilation 
operators $a^\dagger, \tilde{a}^\dagger,  a$,  and  $\tilde{a}$  as 
follows
\begin{eqnarray}
a\vert n,m\rangle &=& \sqrt{n} \vert n-1,m\rangle,\; 
a^\dagger\vert n,m\rangle = \sqrt{n+1} \vert n+1,m\rangle,\\
\tilde{a}\vert n,m\rangle &=&  
\sqrt{m} \vert n,m-1\rangle,\;\tilde{a}^\dagger\vert n,m\rangle 
 = \sqrt{m+1} \vert n,m+1\rangle.
\end{eqnarray}
The operators $a$ and $a^\dagger$  commute  with  $\tilde{a}$  and 
$\tilde{a}^\dagger$.  The number states is given $\vert I\rangle$ in eq(\ref{i}) and
it follows that
\begin{equation}
a\vert I\rangle=\tilde{a}^\dagger \vert I\rangle,\; a^\dagger\vert I\rangle = 
\tilde{a}\vert I\rangle.\label{10}
\end{equation}
By applying $\vert I\rangle$ from the right to the master equation (\ref{jf}) and using 
eq (\ref{10}), one gets
\begin{eqnarray}
\frac{\partial}{\partial t}\vert \rho(t)\rangle = -i\hat{H}\vert \rho\rangle\label{sc}
\end{eqnarray}
where
\begin{eqnarray}
 -i\hat{H}=\left[-iG\left({\cal F}
-\tilde{\cal{F}}^{\dagger}+ {\cal F}^{\dagger}-\tilde{\cal{F}}
\right)
+\frac{\kappa}{2}
\left(2{\cal F}\tilde{\cal{F}}
- \tilde{\cal{F}}\tilde{\cal{F}}^{\dagger}- {\cal F}^{\dagger}{\cal F}
\right)\right]\label{ptww}
\end{eqnarray}
Thus, in TFD the problem of solving master equation is reduced to solving a Schroedinger equation. 
Taking the initial state $\vert \rho(0)\rangle$,
the short term solution is 
\begin{eqnarray}
\vert \rho(t)\rangle  &=&exp\left[
-iGt\left({\cal F}+ {\cal F}^{\dagger}
-(\tilde{\cal{F}}^{\dagger}+\tilde{\cal{F}})
\right)
\right]
\vert \rho(0)\rangle.
\label{1}
\end{eqnarray}
Here the ${\cal F}$ and $ {\cal F}^{\dagger}$ satisfy $SU(1,1)$
commutation relations, similarly for $\tilde{\cal{F}}$,
thus using disentanglement theorem one gets 
\begin{eqnarray}
\vert \rho(t)\rangle \propto exp\left[
-\Gamma_1 {\cal F}^{\dagger}
\right]exp\left[
-\Gamma_2 \tilde{\cal{F}}^{\dagger}
\right]\vert 0\rangle
\end{eqnarray}
 which is an eigenstate of ${\cal{G}}$ and $\tilde{\cal{G}}$ with the eigenvalues $\Gamma_1$ and $\Gamma_2$,
which contains $Gt$, respectively.
By rewriting eq(\ref{ptww}), one gets
\begin{eqnarray}
\frac{\partial}{\partial t}\vert \rho\rangle  &=&
\left[({\cal F}-  \tilde{\cal{F}}^{\dagger}) (-iG+\frac{\kappa}{2} \tilde{\cal{F}})
-(\tilde{\cal{F}}- {\cal F}^{\dagger})(iG+\frac{\kappa}{2}{\cal F})
\right]
\vert \rho\rangle.
\end{eqnarray}
Thus, the steady state solution is given by
\begin{eqnarray}
 \tilde{\cal{F}}\vert \rho\rangle =\frac{2iG}{\kappa}\vert \rho\rangle\label{2}\\
{\cal F}\vert \rho\rangle  =-\frac{2iG}{\kappa}\vert \rho\rangle\label{3}
\end{eqnarray}

If ${\cal F}=ab$ and $ \tilde{\cal{F}}=\tilde{a}\tilde{b}$,
then the short term solution is the Caves-Schumaker state from eq(\ref{1})
\begin{eqnarray}
\vert \rho\rangle  &=&exp\left[
-iGt(ab+a^\dagger b^\dagger-(\tilde{a}^{\dagger}\tilde{b}^{\dagger}+ \tilde{a}\tilde{b}))
\right]
\vert \rho(0)\rangle.
\end{eqnarray}
and the steady state is the  pair coherent state from eq(\ref{2}) and (\ref{3}) given by
\begin{eqnarray}
\tilde{a}\tilde{b}\vert \rho\rangle  =\frac{2iG}{\kappa}\vert \rho\rangle,\\
ab\vert \rho\rangle  =-\frac{2iG}{\kappa}\vert \rho\rangle.
\end{eqnarray}
If ${\cal F}=a^2,$ and $\tilde{\cal{F}}=\tilde{a}^2$, then
the short term is the Yuen state from eq(\ref{1})
\begin{eqnarray}
\vert \rho\rangle  &=&exp\left[
-iGt(a^2+a^{\dagger 2}-(\tilde{a}^{\dagger 2}+ \tilde{a}^2))
\right]
\vert \rho(0)\rangle.
\end{eqnarray}
and steady state is the  Cat state from eq(\ref{2}) and (\ref{3}) given by
\begin{eqnarray}
\tilde{a}^2\vert \rho\rangle  =\frac{2iG}{\kappa}\vert \rho\rangle,\\
a^2\vert \rho\rangle  =-\frac{2iG}{\kappa}\vert \rho\rangle.
\end{eqnarray}
The master equation (\ref{jf}) also generates 
the two boson creation and annihilation operators \cite{vsc1},
for which ${\cal F}=a+\beta a^{\dagger 2}$ and $ \tilde{\cal{F}}=\tilde{a} + 
\beta \tilde{a}^{\dagger 2}$ with the canonical conjugates 
 ${\cal G}_0^\dagger=\frac{a^{\dagger 2}}{2}
\frac{1}{n_a+1}$, ${\cal G}_1^\dagger=\frac{a^{\dagger 2}}{2}\frac{1}{n_a+2}$,
$\tilde{\cal G}_0^\dagger=\frac{\tilde{a}^{\dagger 2}}{2}
\frac{1}{\tilde{n}_a+1}$ and 
$\tilde{\cal G}_1^\dagger=\frac{\tilde{a}^{\dagger 2}}{2}\frac{1}{\tilde{n}_a+2}$, here
$n_a=a^\dagger a$ and $\tilde{n}_a=\tilde{a}^\dagger \tilde{a}$, satisfies 
$[{\cal F},{\cal G}_i^\dagger]=1$ and $[\tilde{\cal F},\tilde{\cal G}_i^\dagger]=1$.

Since the ${\cal F}$ and $ {\cal F}^{\dagger}$ commutes, similarly 
for $\tilde{\cal{F}}$, thus, by using disentanglement theorem, 
the short term solution is given from eq(\ref{1})
and the steady state is given from eq(\ref{2}) and (\ref{3}) are
\begin{eqnarray}
(a+\beta a^{\dagger 2})\vert \rho\rangle & =&\frac{2iG}{\kappa}\vert \rho\rangle,\\
(\tilde{a} + \beta \tilde{a}^{\dagger 2})\vert \rho\rangle &=&0-\frac{2iG}{\kappa}\vert \rho\rangle.
\end{eqnarray}
Another two boson creation and annihilation operators \cite{vsc1}, 
for which ${\cal F}=ab+\beta a^{\dagger}b^{\dagger}$ and 
$ \tilde{\cal{F}}=\tilde{a} \tilde{b}+ 
\beta \tilde{a}^{\dagger }\tilde{b}^{\dagger }$ with the 
canonical conjugates  ${\cal G}_0^\dagger=\frac{a^{\dagger}b^{\dagger}}{2}
\frac{1}{n_b+1}$, ${\cal G}_1^\dagger=\frac{a^{\dagger}b^{\dagger}}{2}\frac{1}{n_a+1}$, 
$\tilde{\cal G}_0^\dagger=\frac{\tilde{a}^{\dagger}\tilde{b}^{\dagger}}{2}
\frac{1}{\tilde{n}_b+1}$ and 
$\tilde{\cal G}_1^\dagger=\frac{\tilde{a}^{\dagger}\tilde{b}^{\dagger}}{2}\frac{1}{\tilde{n}_a+1}$, here
$n_a=a^\dagger a$, $n_b=b^\dagger b$,
$\tilde{n}_a=\tilde{a}^\dagger \tilde{a}$ and $\tilde{n}_b=\tilde{b}^\dagger \tilde{b}$, satisfies 
$[{\cal F},{\cal G}_i^\dagger]=1$ and $[\tilde{\cal F},\tilde{\cal G}_i^\dagger]=1$.
Again here the ${\cal F}$ and $ {\cal F}^{\dagger}$ commutes, similarly for $\tilde{\cal{F}}$, thus, the short term solution is given from eq(\ref{1})
and the steady state is given from eq(\ref{2}) and (\ref{3}) are
\begin{eqnarray}
(ab+\beta a^{\dagger}b^{\dagger})\vert \rho\rangle  &=&\frac{2iG}{\kappa}\vert \rho\rangle,\\
(\tilde{a} \tilde{b}+ \beta \tilde{a}^{\dagger }\tilde{b}^{\dagger })\vert \rho\rangle 
&=&-\frac{2iG}{\kappa}\vert \rho\rangle.
\end{eqnarray}

To conclude we have constructed a general master equation
for Janus Faced commutation relations $[{\cal F},{\cal G}^\dagger]=1$
and shown that the eigenstate ${\cal G}$ occurs as a short term solution
and the eigenstate ${\cal F}$ occurs as a steady state solution. We have 
also shown that the master equation works for all the Janus Faced coherent state
operators.
\section{Acknowledgements}
The authors would like to thank M.S.Sriram for his nice comments and encouragement.

\end{document}